\documentclass[aps,groupedaddress,superscriptaddress,amssymb,amsmath,amsbsy,amsfonts,twocolumn,pra]{revtex4}

\usepackage{amssymb,amsmath,amsthm,color,graphicx,times,graphicx}
\usepackage{hyperref}
\usepackage{enumerate}
\usepackage{subfigure}
\usepackage{dsfont}
\usepackage{times}
\usepackage{txfonts}
\usepackage{bbold}

\newcommand*{\cl}[1]{{\mathcal{#1}}}
\newcommand*{\bb}[1]{{\mathbb{#1}}}
\newcommand{\ket}[1]{\left|#1\right>}

\newcommand{\proj}[2]{| #1 \rangle\!\langle #2 |}
\newcommand*{\tn}[1]{{\textnormal{#1}}}

\newcommand{\T}{\mbox{$\textnormal{Tr}$}}

\usepackage{color}

\theoremstyle{plain}

\newtheorem{theorem}{Theorem}
\newtheorem{lemma}[theorem]{Lemma}
\newtheorem{remark}{Remark}

\theoremstyle{definition}

\date{\today}

\begin{document}

\title{Universal upper bounds for Gaussian information capacity}

\author{Kabgyun Jeong}
\email{kgjeong6@snu.ac.kr}
\affiliation{IMDARC, Department of Mathematical Sciences, Seoul National University, Seoul 08826, Korea}
\affiliation{School of Computational Sciences, Korea Institute for Advanced Study, Seoul 02455, Korea}

\author{Hun Hee Lee} 
\affiliation{IMDARC, Department of Mathematical Sciences, Seoul National University, Seoul 08826, Korea}

\author{Youngrong Lim} 
\affiliation{School of Computational Sciences, Korea Institute for Advanced Study, Seoul 02455, Korea}

\begin{abstract}
The most natural way to describe an information-carrying system containing a specific noise is an additive white Gaussian-noise (AWGN) channel. In bosonic quantum systems (especially the Gaussian case), although the classical information capacity for a phase-insensitive and thermal-noise channel is additive based on a proof of the minimum output entropy conjecture, several open questions remain. By generalizing the Gaussian noise model from thermal noise to general Gaussian noise, we rigorously revisit and calculate these strong upper bounds on the information capacity for single-mode with general Gaussian-noise channels. In this study, we use the quantum entropy power inequality (QEPI) approach. This framework gives a new formula for finding upper bounds on the information capacity of bosonic Gaussian channels.
\end{abstract}

\maketitle

\section{Introduction}
A central goal in information theory is to find the channel capacity of a communication system~\cite{S48}. In quantum Shannon theory, a quantum analogue of information theory, one of most important and challenging tasks is to determine the channel capacity of a quantum communication system~\cite{NC00,W13,H06,HW01}. The channel capacity of a communication channel is generally defined as the maximal information transmission rate at which certain information can be transmitted reliably through an electromagnetic channel within vanishing errors. A communication system can be characterized by a quantum optical channel modeled by an input bosonic quantum state transformed into an output bosonic state with its external thermal noise (via Gaussian unitaries); this is called a thermal-noise channel. Herein, we restrict ourselves to a quantum channel described only by Gaussian operations over bosonic Gaussian states and a bosonic Gaussian noise, which is a quantum analogue corresponding to the classical additive white Gaussian-noise (AWGN) channel. In general, it is not easy to determine the channel capacity of a given quantum channel in quantum Shannon theory~\cite{H06}, and it is almost impossible to obtain its ultimate channel capacity or \emph{information} capacity when quantum entanglement is involved. In fact, most channel capacities for a quantum channel are known to be nonadditive~\cite{H09,SY08,LWZG09}.

Recently, proving a Gaussian minimum output entropy conjecture~\cite{GGCH14,MGH14} over a quantum channel has been suggested as one way to obtain a tight upper bound on the Gaussian information capacity for a thermal-noise channel. This implies that the Gaussian information capacity of a thermal-noise channel is additive because  it saturates to the well-known Holevo capacity. By contrast, a universal upper bound on capacity can be obtained by exploiting a new notion of quantum entropy power inequality (QEPI) in the Gaussian regime~\cite{KS13,KS13+}. Owing to  the potential role of Gaussian entanglement, such as the nonadditivity of the Gaussian information capacity, we must carefully consider the upper bounds derived using the QEPI. We still believe that there exists some additivity violation for bosonic Gaussian-noise channels if any entangled (or squeezed) encodings are possible. If not, Gaussian channels may be useless from the viewpoint of increasing the channel capacity for Gaussian communications via quantum entanglement. This is why we carefully consider the upper bounds on the information capacity through the QEPI in the Gaussian regime.

Entropy power inequality (EPI) is an important tool in information theory. Shannon~\cite{S48} proposed the classical version of EPI, and K\"{o}nig and Smith~\cite{KS14} proposed QEPI. Many seminal papers have provided the proofs for EPI~\cite{L78,DCT91,B75,BL76,R11,S59,B65} and QEPI~\cite{KS14,PMG14,PMG14+,PMLG15,ADO16,CLL16,K15,JLJ18,PT17} and reported their implications for bosonic Gaussian information capacities ~\cite{KS13,KS13+}. We briefly introduce Gaussian QEPI. Let $\rho_{X_1}$ and $\rho_{X_2}$ be two $D$-mode independent Gaussian input signals and assume that the signals interact at the beamsplitter with a mixing parameter $\tau\in[0,1]$. Then,  the entropy of the resulting Gaussian output signal is lower-bounded as follows~\cite{KS14,PMG14}:
\begin{equation}
\exp\left(\frac{S(\rho_{X_1}\boxplus_\tau\rho_{X_2})}{D}\right)\ge{\tau}\exp\left(\frac{S(\rho_{X_1})}{D}\right)+{(1-\tau)}\exp\left(\frac{S(\rho_{X_2})}{D}\right), \label{eq:GQEPI}
\end{equation}
where $\boxplus_\tau$ denotes the beamsplitting operation on the independent two input signal states under mixing parameters $\tau$ and $S(\rho)=-\T(\rho\log\rho)$, the von Neumann entropy in nats. Similarly, Eq.~(\ref{eq:GQEPI}) has the entropic form $S(\rho_{X_1}\boxplus_\tau\rho_{X_2})\ge\tau S(\rho_{X_1})+(1-\tau)S(\rho_{X_2})$ from the concavity of the entropic function; however, there is no explicit proof for this equivalence.

Next, we investigate a bosonic Gaussian channel including \emph{general} Gaussian noise, for example, environmental noise induced from a convex combination of Gaussian states such as coherent~\cite{K05} and squeezed states~\cite{JKL15,JL16} via general Gaussian unitaries. Here, we explicitly calculate the universal upper bounds of information capacity for general Gaussian-noise channels and derive a new formula. Our case is broader in the sense of general Gaussian noise involving squeezed states up to any phase rotation. Furthermore, we suggest an exact formula to restrict universal bounds for the Gaussian information capacity of a bosonic Gaussian channel, where the lower bound is naturally obtained from the well-known Holevo-Schumacher-Westmoreland theorem~\cite{H98,SW97} and the upper bounds are obtained from Smith and K\"{o}nig's works~\cite{KS13,KS13+} and the present study. In addition, we extend our result to a quantum amplifier, whose interaction is represented by an amplification involving squeezing operations. The corresponding QEPI can be written as~\cite{PMG14}
\begin{equation}
\exp\left(\frac{S(\rho_{X_1}\boxplus_\kappa\rho_{X_2})}{D}\right)\ge{\kappa}\exp\left(\frac{S(\rho_{X_1})}{D}\right)+{(\kappa-1)}\exp\left(\frac{S(\rho_{X_2})}{D}\right), \label{amp}
\end{equation}
where $\boxplus_\kappa$ is the amplifying operation with the parameter $\kappa>1$. Finally we make a comparison between our result and an upper bound of the Gaussian information capacity, i.e., the classical capacity restricted for Gaussian input states, of the phase-sensitive Gaussian channels~\cite{SKPC16}.

\begin{figure}
\centering
\includegraphics[width=\columnwidth]{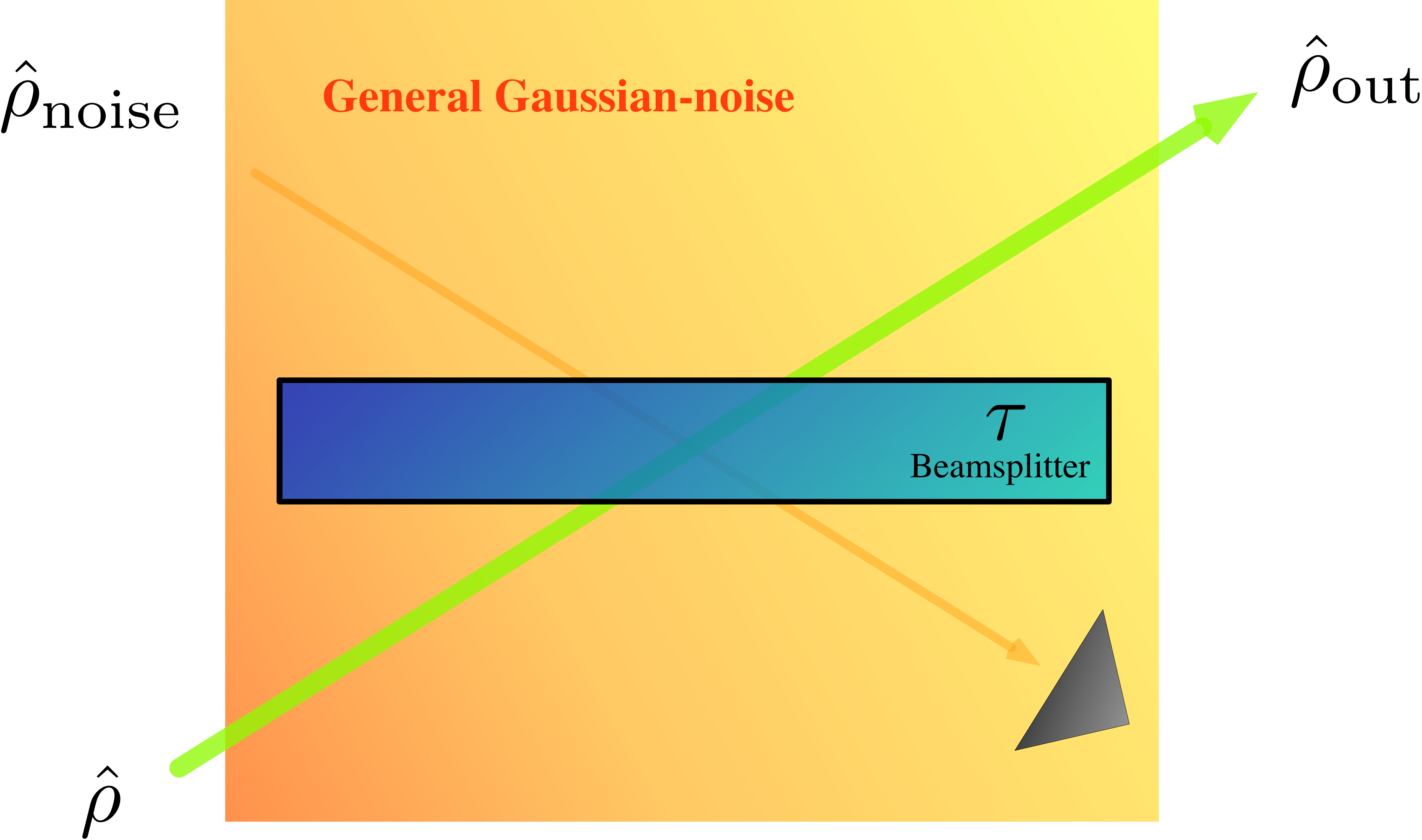}
\caption{(Color online) {\bf A model for a general Gaussian-noise channel through a $\tau$-beamsplitting operation.} For given Gaussian channel input state $\hat{\rho}$, the general Gaussian-noise channel is described by an environmental input state $\hat{\rho}_\tn{noise}$ that  is given by the thermal noise $\hat{\rho}_\tn{th}$ or general Gaussian noise $\hat{\rho}_G$. A Gaussian channel output state of the Gaussian-noise channel is denoted by $\hat{\rho}_\tn{out}$, and the black triangle indicates the partial trace of the second (or noise) mode after the $\tau$-beamsplitting operation.}
\label{fig:BS}
\end{figure}

\section{Model of general Gaussian-noise channel}

In the Gaussian quantum regime~\cite{WP+12}, a bosonic Gaussian-noise channel is usually described as follows:
\begin{align}
\hat{c}
&=\sqrt{\tau}\hat{a}+\left\{ \begin{array}{ll} 
\sqrt{1-\tau}\hat{b},\ & \tau\in[0,1];\\ & \\
\sqrt{\tau-1}\hat{b}^\dag,\ & \tau>1,
\end{array}\right.
\end{align}
where $\hat{a}, \hat{b}$ represent input bosonic (annihilation) field operators and $\hat{c}$ represents an output mode satisfying a specific canonical commutation relation. In this case, it is known that the Gaussian-noise channel can be decomposed into two types of quantum channels given by a lossy and an amplifying channel with $\tau\in[0,1]$ and $\tau>1$, respectively. If $\tau<1$ with environmental mode $\hat{b}$ take the vacuum state, the channel is called a pure lossy channel. The most important class of quantum channel of this type is called a thermal-noise channel $\cl{N}_{\tau,\bar{N}_E}(\hat{\rho})$, in which the input state $\hat{\rho}$ is a bosonic Gaussian state with mean photon number $\bar{N}$. In the state representation via the mixing operation given by $\tau$-\emph{beamsplitter}, the thermal-noise channel is described in the form of Stinespring dilation as
\begin{align}
\cl{N}_{\tau,\bar{N}_E}(\hat{\rho})&=\T_{E}\left[U_{\tau}(\hat{\rho}\otimes\hat{\rho}_\tn{th})U_{\tau}^\dagger\right]:=\hat{\rho}_\tn{out},
\end{align}
where $\hat{\rho}_\tn{th}$ is the thermal state with a fixed mean photon number $\bar{N}_E$, and $U_{\tau}$, the $\tau$-beamsplitter operation over the system and environment $E$. The thermal state is naturally defined by
\begin{align}
\hat{\rho}_\tn{th}(\bar{N}_E)
&=\sum_{N=0}^\infty\frac{\bar{N}_E^N}{(\bar{N}_E+1)^{N+1}}\proj{N}{N},
\end{align}
where $\{\ket{N}\}_{N=0}^\infty$ denotes a set of number (or Fock) basis. Under a partial trace over environmental system $E$ (i.e., thermal-noise mode), Fig.~\ref{fig:BS} precisely shows the situation in which the channel's output is given by a Gaussian state $\hat{\rho}_\tn{out}$. In Fig.~\ref{fig:BS}, $\hat{\rho}_\tn{noise}$ is given by the thermal noise $\hat{\rho}_\tn{th}$ or the general Gaussian noise $\hat{\rho}_G$ including squeezed states up to phase rotations. For any multimode channel case, the thermal noise acts independently at each mode of Gaussian channels corresponding to each input signal because an environmental subsystem can be considered a generic big-bath as in traditional thermodynamics.

Next, we consider the general Gaussian-noise model in the covariance matrix framework. As described in the Supplemental Material, we can easily find a (single-mode) covariance matrix on general Gaussian noise such that the mean photon number is $\langle\hat{b}^\dag\hat{b}\rangle:=\bar{N}_E=(\tfrac{1}{2}\T V_G-1)/2$. For any real parameters $\theta,r$ and $\phi\in\mathbb{R}$,
\begin{equation} \label{eq:gcov}
V_G=\Sigma_TV_\tn{th}\Sigma_T^{\top}=(2\bar{N}_\tn{th}+1)O(\theta)T(2r)O^{\top}(\theta),
\end{equation}
where any single-mode Gaussian symplectic matrix (or transform) can always be decomposed as $\Sigma_T=O(\theta)T(r)O(\phi)$ if $T$ is a symplectic matrix under a squeezing parameter $r$ and a $2\times2$ matrix $O(\cdot)$ denotes the phase-rotation operator on the symplectic space~\cite{WP+12}. $V_\tn{th}$ is the covariance matrix for the thermal state, and it is given by
$V_\tn{th}=V(\hat{\rho}_\tn{th})=(2\bar{N}_\tn{th}+1)\tn{\bf I}$.
One conceptual example of $V_G$ above is the covariance matrix of squeezed states with zero-mean value. Specifically, if we set $\theta=0$ and $\bar{N}_\tn{th}$ is the thermal photon number of the environment, then
\begin{equation}
V_\tn{sq}:=V(\hat{\rho}_\tn{sq})=(2\bar{N}_\tn{th}+1)
\left( \begin{array}{cc}
e^{-2r} & 0 \\ & \\
0 & e^{2r} \end{array}\right).
\end{equation}
This implies that general Gaussian noise generated by displacements and squeezing operations can always be obtained via conjugated Gaussian unitaries over the thermal state. For a squeezed-thermal environmental state, the relation between the mean photon number $\bar{N}_E$ and the thermal photon number $\bar{N}_\tn{th}$ of the environment is generally given by $2\bar{N}_E+1=(2\bar{N}_\tn{th}+1)\cosh 2r$.

Next, we consider a determinant for the covariance matrices for general Gaussian noise. From the covariance matrices with the form $V_\tn{th}$ and $V_\tn{sq}$, we can prove that
$\det V_\tn{th}=(2\bar{N}_\tn{th}+1)^2$ and $\det V_\tn{sq}=(2\bar{N}_\tn{th}+1)^2$, respectively. Thus, we find the determinant of $V_G$ of $\hat{\rho}_G$ as (see more details in Supplemental Material)
\begin{equation}
\det V_G=(2\bar{N}_E+1)^2\;\;\tn{where}\;\bar{N}_E=\bar{N}_{\tn{th}},
\end{equation}
which mainly contributes to constructing a general formula for universal upper bounds on Gaussian information capacity.

\section{Information capacity and Gaussian minimum output entropy conjecture}
As mentioned above, it is difficult to determine the information capacity of a quantum channel. However, although it is still restricted to explicitly calculating the capacity, some breakthroughs have been made on calculating the information capacity of bosonic Gaussian-noise channels (i.e., thermal-noise channels). Next, we briefly review known results on the quantum channel capacity of bosonic Gaussian-noise channels.

We consider a Gaussian input signal for some Gaussian quantum channels with mean photon number $\bar{N}(:=\langle\hat{a}^{\dag}\hat{a}\rangle)$, and the channel is only restricted to a single-mode case. In the absence of environmental noise, that is, bosonic Gaussian-noiseless channel $\cl{N}_0$ (i.e., $\tau=0$), the information capacity (in nats) is given by $C(\cl{N}_0,\bar{N})=g(\bar{N})$~\cite{YO93,CD94}, where the entropic function is defined by $g(x):=(1+x)\log(1+x)-x\log x$. Furthermore, it is well-known that $C(\cl{N}_{\tau,0},\bar{N})=g(\tau\bar{N})$ for a pure lossy Gaussian channel $\cl{N}_{\tau,0}$. If a bosonic Gaussian-noise channel involves any general Gaussian noise, that is, general Gaussian-noise channel $\cl{N}_{\tau,\bar{N}_E}$, we consider how we can determine or bound the information capacity of a Gaussian quantum channel. In this study, we claim that for any input Gaussian state with mean photon number $\bar{N}$,
\begin{equation}
\chi(\cl{N}_{\tau,\bar{N}_E},\bar{N})\le C(\cl{N}_{\tau,\bar{N}_E},\bar{N})\le C_{\max}(\cl{N}_{\tau,\bar{N}_E},\bar{N}),
\end{equation}
where $\chi$ is the well-known Holevo capacity for the Gaussian-noise channel by exploiting coherent-state encodings~\cite{GLMS03}. The maximal information capacity $C_{\max}$ is bounded below by the QEPI with the $\tau$-beamsplitter operation. In the thermal-noise case, if we take the input ensemble as coherent states $\{q_i,\alpha_i\}$ (more explicitly, $q_i=\tfrac{1}{\pi\bar{N}}\exp({-\tfrac{|\alpha_i|^2}{\bar{N}}})$ and $\alpha_i=\proj{\alpha_i}{\alpha_i}$) having average state $\bar{\alpha}=\int\tfrac{1}{\pi\bar{N}}\exp({-\tfrac{|\alpha_i|^2}{\bar{N}}})\proj{\alpha_i}{\alpha_i}d\alpha_i$, the Holevo capacity with the coherent-state encodings of the channel $\cl{N}_{\tau,\bar{N}_E}$ are given by
\begin{equation}
\chi(\cl{N}_{\tau,\bar{N}_E},\bar{N})=g(\tau\bar{N}+(1-\tau)\bar{N}_E)-g((1-\tau)\bar{N}_E).
\end{equation}
From the regularization of the Holevo capacity, we can naturally define the Gaussian information capacity as
$C(\cl{N}_{\tau,\bar{N}_E},\bar{N})=\lim_{n\to\infty}\frac{1}{n}\chi(\cl{N}_{\tau,\bar{N}_E}^{\otimes n},n\bar{N})$, which results in the most famous conjecture in quantum Shannon theory, that is, $C(\cl{N}_{\tau,\bar{N}_E},\bar{N})=\chi(\cl{N}_{\tau,\bar{N}_E},\bar{N})$. Recent studies on the conjecture have shown that it is true in a bosonic and phase-insensitive Gaussian-noise channel~\cite{GGCH14,MGH14}. The proof is given by solving the Gaussian minimum output entropy conjecture~\cite{GGLMS04,GGL+04}: for any Gaussian state $\hat{\rho}$ with mean photon number $\bar{N}$,
\begin{equation} \label{MOE}
S_G^{\min}(\cl{N}_{\tau,\bar{N}_E},\bar{N}):=\lim_{n\to\infty}\min_{\hat{\rho}}\frac{1}{n}S\left(\cl{N}_{\tau,\bar{N}_E}^{\otimes n},\hat{\rho}_n\right)=g((1-\tau)\bar{N}_E),
\end{equation}
where $\hat{\rho}_n$ denotes any $n$-fold input states with a potential quantum entanglement.
We note that the maximal information capacity of this channel in terms of the minimum output entropy statement is given by $C_{\max}(\cl{N}_{\tau,\bar{N}_E},\bar{N})=g(\tau\bar{N}+(1-\tau)\bar{N}_E)-S_G^{\min}(\cl{N}_{\tau,\bar{N}_E},\bar{N})$, where the first term is obtained from the maximal entropy achieved in the Gaussian states~\cite{WGC06}. Thus, the information capacity saturates to the Holevo capacity under coherent-state encodings.

However, there is more to the Gaussian information capacity, because they only consider the coherent-state encodings at the input bosonic Gaussian states over a single-mode scenario. If a Gaussian channel involves a phase-sensitive \emph{squeezing} element, we cannot confirm the saturation on the capacity as in Eq.~(\ref{MOE})~\cite{PLM12}. Therefore, we must consider the QEPI in the Gaussian regime to find universal bounds for the Gaussian information capacity.

\begin{figure*}
\centering
\includegraphics[width=\linewidth]{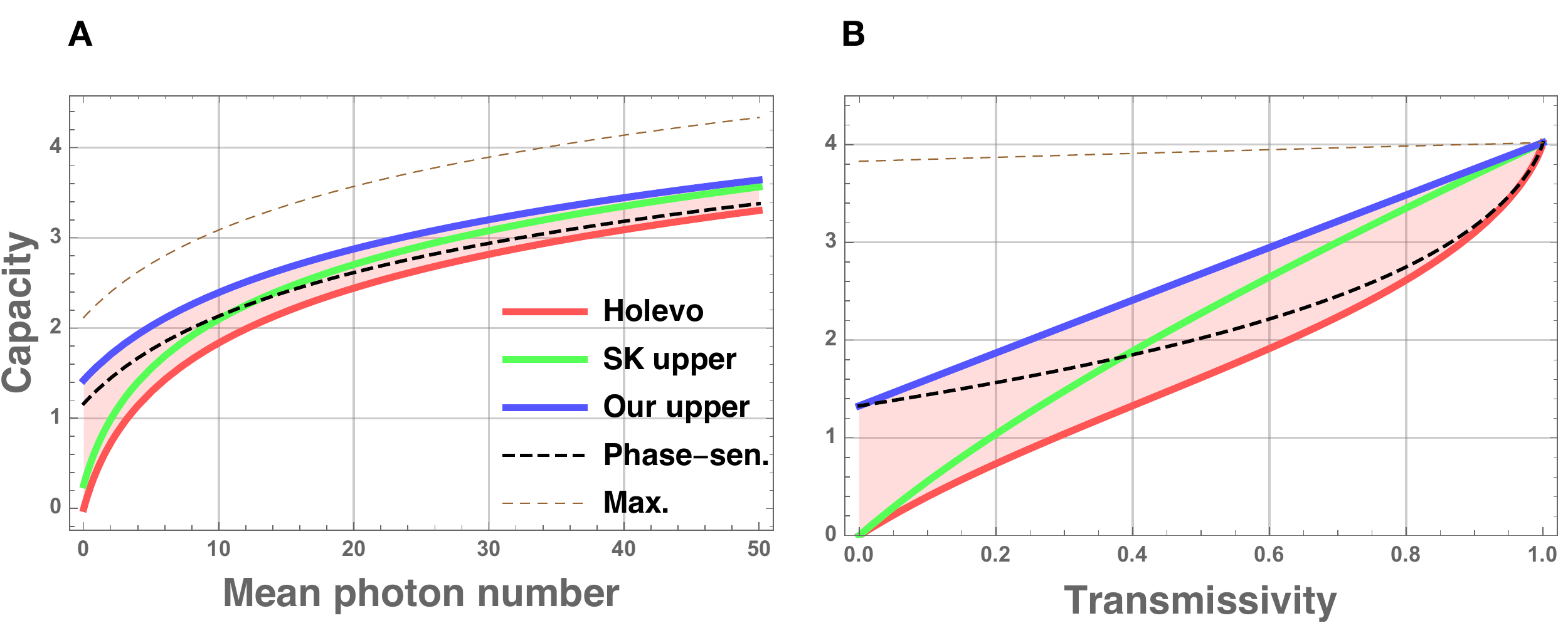}
\caption{(Color online) {\bf Upper bounds on the information capacity of a general Gaussian-noise channel $C(\cl{N}_{\tau,\langle V_\tn{th}\rangle},\bar{N})$ in nats.} ({\textbf A}) Plots of information capacity bounds for transmissivity $\tau=\frac{1}{2}$ and environmental mean photon number $\bar{N}_E=1$ and squeezing parameter $r=1$. ({\textbf B}) Plots of information capacity bounds for input mean photon number $\bar{N}=20$, $\bar{N}_E=4$, and $r=1$. `Phase-sen.' denotes the upper bound of Gaussian information capacity for phase-sensitive channel given in Ref.~\cite{SKPC16}. When $r=0$ (squeezing-free noise), our upper bound is equal to `SK upper' bound, and `Phase-sen.' bound is equal to `Holevo' bound as expected.}
\label{fig:Upper}
\end{figure*}

\section{Gaussian information capacity of general Gaussian-noise channel}
We recall that the general Gaussian-noise model is represented by the single-mode covariance matrix $V_G$ of $\hat{\rho}_\tn{noise}$ as in Eq.~(\ref{eq:gcov}) with zero-mean and the determinant $\det{V_G}=(2\bar{N}_E+1)^2$. Our main result for the Gaussian information capacity is as follows. Let $\cl{N}_{\tau,\langle V_G\rangle}$ denote the general Gaussian-noise channel. Then, a universal upper bound for the Gaussian information capacity in nats of the channel is determined as ($\forall~\tau\in(0,1)$)
\begin{equation}
C(\cl{N}_{\tau,\langle V_G\rangle},\bar{N})\le g\left(\tau\bar{N}+(1-\tau)\bar{N}_E\right)-(1-\tau)g\left(\tfrac{1}{2}(\sqrt{{\det V_G}}-1)\right),
\end{equation}
where $\bar{N}=\langle\hat{a}^\dag\hat{a}\rangle$ is the mean photon number for the input single mode $\hat{a}$, and $\langle V_G\rangle:=\langle\hat{b}^\dag\hat{b}\rangle$ is the mean photon number of the environmental noise $\hat{\rho}_\tn{noise}$. The proof of this formula is directly given by the QEPI~\cite{KS14,PMG14} and the abovementioned bound of the maximal information capacity. First, QEPI in the Gaussian regime for the multimode gives
\begin{equation}
S\left(\cl{N}_{\tau,\langle V_G\rangle}^{\otimes n},\hat{\rho}_n\right)\ge n(1-\tau)g\left(\tfrac{1}{2}(\sqrt{{\det V_G}}-1)\right),
\end{equation}
where $\hat{\rho}_n\simeq\hat{\rho}^{\otimes n}$ represents the Gaussian input encoded states over the multimode bosonic channels (they can potentially be \emph{entangled} encodings), and for convenience, we consider that the Gaussian input signal $\hat{\rho}$ has mean photon number $\bar{N}$. Furthermore, note that $\hat{\rho}$ and $\hat{\rho}_\tn{noise}$ are initially independent before the mixing operation with the $\tau$-beamsplitter.
Second, it is straightforwardly given by the maximal information capacity $C_{\max}(\cl{N}_{\tau,\langle V_G\rangle},\hat{\rho})=g(\tau\bar{N}+(1-\tau)\bar{N}_E)-\lim_{n\to\infty}\min_{\hat{\rho}}\frac{1}{n}S\left(\cl{N}_{\tau,\langle V_G\rangle}^{\otimes n},\hat{\rho}_n\right)$. These two arguments complete the proof for the universal upper bounds of the Gaussian information capacity.

We can easily check that this formula exactly reproduces the information capacity's upper bound for a thermal-noise channel~\cite{KS13} as follows:
\begin{equation}
C(\cl{N}_{\tau,\langle V_\tn{th}\rangle},\bar{N})\le g\left(\tau\bar{N}+(1-\tau)\bar{N}_E\right)-(1-\tau)g\left(\bar{N}_E\right),
\end{equation}
because $\tfrac{1}{2}(\sqrt{{\det V_\tn{th}}}-1)=\tfrac{1}{2}(2\bar{N}_\tn{th}+1)-\tfrac{1}{2}$ and $\bar{N}_E=\bar{N}_\tn{th}$ for the thermal noise.
 
Similarly, from the QEPI Eq.~(\ref{amp}), we can obtain an upper bound of amplifier with general Gaussian noise as
\begin{align}
C(\cl{N}_{\kappa,\langle V_G\rangle},\bar{N}) \le \nonumber&~ g\left(\kappa\bar{N}+(\kappa-1)(\bar{N}_E+1)\right)\\&-\frac{\kappa-1}{2\kappa-1}g\left(\tfrac{1}{2}(\sqrt{{\det V_G-1}})\right).
\end{align}
Then it gives the information capacity's upper bound for the thermal-amplifier channel as follows:
\begin{equation}
C(\cl{N}_{\kappa,\langle V_\tn{th}\rangle},\bar{N}) \le g\left(\kappa\bar{N}+(\kappa-1)(\bar{N}_E+1)\right)-\frac{\kappa-1}{2\kappa-1}g\left(\bar{N}_E\right).
\end{equation}

Finally, we compare our result with a Gaussian information capacity's upper bound for phase-sensitive channel~(Eq. (29) in~\cite{SKPC16}) in Fig.~\ref{fig:Upper}, and also we make plots for the comparison with the Holevo bound and the maximal capacity bound. It seems that the upper bound for the Gaussian information capacity for the phase-sensitive case is tighter than ours, but it doesn't mean our upper bound is indeed weaker owing to $C^G\le C$. When $r=0$ (thermal noise), this Gaussian information capacity is equal to `Holevo' bound, which means that we can obtain the exact formula of classical capacity of the channels. This is expected because of the fact that Gaussian optimizer conjecture was proved for the thermal-noise channel. Also for the case of thermal-noise channel, our upper bound is saturated on the `Smith-K\"{o}nig (SK) upper' bound as expected, however it gives new bounds for general Gaussian-noise one including a squeezing element.

\section{Discussions}
In this study, we have described fundamental and universal upper bounds on Gaussian information capacity of bosonic general Gaussian-noise channels, which include thermal-noise channels. The general Gaussian-noise channel naturally involves both coherent states as well as squeezed states up to the phase rotations on the environmental subsystem; in this sense, it can be considered a general case of the Gaussian-noise model. By exploiting the fact that the determinants of the covariance matrices of Gaussian states are essentially equivalent, we suggest a new formula for the information capacity of a Gaussian quantum regime.

In the Gaussian quantum regime, although the additivity of the classical capacity for phase-insensitive thermal-noise channels is true under the Gaussian minimum output entropy conjecture (i.e., channel capacity is additive), several unsolved problems remain. Herein, we rigorously calculate the upper bounds on the classical information capacity of single-mode general Gaussian-noise channels (including squeezing components or potential quantum entanglement) via the QEPI. In this framework, we have found a new formula for finding the upper bounds for the information capacity of bosonic Gaussian-noise channels.

Our study may be closely related to the quantum entropy-photon-number inequality (EPNI) first proposed by Guha et al.~\cite{G04,GES08}, and it leads to explicit calculations for various Gaussian channel capacities. If the EPNI holds in the Gaussian regime, our understanding of the Gaussian information capacity will improve and may result in various potential applications.

Finally, we mention that our result can be derived from Lemma 1A in Ref.~\cite{HK18} and the concavity property of the logarithmic function. Explicitly, in the case of $S(\rho_\tn{noise}):=g(\frac{1}{2}\sqrt{\det V_G}-1)$,
\begin{align*}
C(\cl{N}_{\tau,\langle V_G\rangle},\bar{N})&\le g\left(\tau\bar{N}+(1-\tau)\bar{N}_E\right)-\log\left(\tau+(1-\tau)e^{S(\rho_\tn{noise})}\right) \\
&\le g\left(\tau\bar{N}+(1-\tau)\bar{N}_E\right)-(1-\tau)S(\rho_\tn{noise}).
\end{align*}
However, we used a linear Gaussian version not in non-Gaussian one in the framework of the quantum entropy power inequality.

\section{Acknowledgments}
We are grateful to Dong Pyo Chi, Changhun Oh, and Wookyum Kim 
for valuable comments. This work was supported by the National Research Foundation of Korea (NRF) through a grant funded by the Ministry of Science and ICT (NRF-2017R1E1A1A03070510 \& NRF-2017R1A5A1015626) and Ministry of Education (NRF-2018R1D1A1B07047512).
 

\newpage
\onecolumngrid
\appendix

\setcounter{equation}{0}
\setcounter{figure}{0}
\setcounter{table}{0}
\renewcommand{\theequation}{S\arabic{equation}}
\renewcommand{\thefigure}{S\arabic{figure}}
\renewcommand{\bibnumfmt}[1]{[S#1]}
\renewcommand{\citenumfont}[1]{S#1}

\begin{center}
{\large $\langle$ Supplemental Material $\rangle$}
\vspace{0.7 EM}

{\large\bf Universal upper bounds for Gaussian information capacity}
\end{center}

\section{Covariance matrices and its determinants} \label{app:cov}
For every formal symplectic matrix $T\in\cl{T}$, the set of symplectic matrices satisfies $\cl{T}\Omega\cl{T}^\top=\Omega$ where $\Omega:=\left( \begin{array}{cc}
0 & 1 \\
-1 & 0 \end{array}\right)$ is called as symplectic form. If we choose a symplectic matrix as $T=\left( \begin{array}{cc}
e^{-r} & 0 \\
0 & e^{r} \end{array}\right)$, i.e., the squeezing operator in which $r$ is a squeezing parameter and $\forall~\theta\in\mathbb{R}$, then we have
\begin{align}
T\Omega T^\top&=
\left( \begin{array}{cc}
e^{-r} & 0 \\ & \\
0 & e^{r} \end{array}\right)
\left( \begin{array}{cc}
0 & 1 \\ & \\
-1 & 0 \end{array}\right)
\left( \begin{array}{cc}
e^{-r} & 0 \\ & \\
0 & e^{r} \end{array}\right)
=\left( \begin{array}{cc}
0 & 1 \\ & \\
-1 & 0 \end{array}\right)=\Omega, \\
TT^\top&=
\left( \begin{array}{cc}
e^{-r} & 0 \\ & \\
0 & e^{r} \end{array}\right)
\left( \begin{array}{cc}
e^{-r} & 0 \\ & \\
0 & e^{r} \end{array}\right)
=\left( \begin{array}{cc}
e^{-2r} & 0 \\ & \\
0 & e^{2r} \end{array}\right),
\end{align}
and also notice that
\begin{align}
O(\theta)O(\theta)^\top=
\left( \begin{array}{cc}
\cos\theta & \sin\theta \\ & \\
-\sin\theta & \cos\theta \end{array}\right)
\left( \begin{array}{cc}
\cos\theta & -\sin\theta \\ & \\
\sin\theta & \cos\theta \end{array}\right)=\tn{\bf I}.
\end{align}

Now let us consider a $2\times2$ symplectic matrix such that $T=O(\theta)T(r)O(\phi)\in\cl{T}$ where $O(\cdot)$ is the phase-rotation operator above and $T(r)$ the squeezing operator with a squeezing parameter $r$. Then, for all $\theta,\phi$, and $r\in\bb{R}$, there exist a single-mode symplectic transformation $S_T$ such that 
\begin{align}
V_G&=\Sigma_TV_\tn{th}\Sigma_T^\top \\
&=O(\theta)T(r)O(\phi)(2\bar{N}+1)\tn{\bf I} O^\top(\phi)T^\top(r)O^\top(\theta) \\
&=(2\bar{N}+1)O(\theta)T(2r)O^\top(\theta),
\end{align}
where we fix the mean photon number as $\bar{N}$ and $V_\tn{th}:=(2\bar{N}+1)\tn{\bf I}$ is a covariance matrix (CvM) of the thermal state. We call the matrix $V_G$ as CvM for a (single-mode) general Gaussian state. Note that $V(\hat{\rho}_\tn{0}):=V_\tn{0}=\tn{\bf I}$ for the vacuum state, which has its minimal value. This process is well-known as \emph{singular value decomposition} on Gaussian state and the determinant of the covariance matrix for the thermal state is give by
\begin{equation}
\det V_\tn{th}=(2\bar{N}+1)^2.
\end{equation}
By expanding the thermal state to general Gaussian one, easily we can obtain a general single-mode CvM for the general Gaussian state (or noise) as 
\begin{align}
V_G&=(2\bar{N}+1)O(\theta)T(2r)O(\theta)^\top \\
&=(2\bar{N}+1)\left( \begin{array}{cc}
\cos\theta & \sin\theta \\ & \\
-\sin\theta & \cos\theta \end{array}\right)
\left( \begin{array}{cc}
e^{-2r} & 0 \\ & \\
0 & e^{2r} \end{array}\right)
\left( \begin{array}{cc}
\cos\theta & -\sin\theta \\ & \\
\sin\theta & \cos\theta \end{array}\right) \\
&=(2\bar{N}+1)\left( \begin{array}{cc}
\cos\theta e^{-2r} & \sin\theta e^{2r} \\ & \\
-\sin\theta e^{-2r} & \cos\theta e^{2r} \end{array}\right)
\left( \begin{array}{cc}
\cos\theta & -\sin\theta \\ & \\
\sin\theta & \cos\theta \end{array}\right) \\
&=(2\bar{N}+1)\left( \begin{array}{cc}
\cos^2\theta e^{-2r}+\sin^2\theta e^{2r} & -\cos\theta\sin\theta e^{-2r}+\cos\theta\sin\theta e^{2r} \\ & \\
-\cos\theta\sin\theta e^{-2r}+\cos\theta\sin\theta e^{2r} & \sin^2\theta e^{-2r}+\cos^2\theta e^{2r} \end{array}\right),
\end{align}
where its determinant is straightforwardly given by following form:
\begin{align}
\det V_\tn{G}&=\left|(2\bar{N}+1)\left( \begin{array}{cc}
\cos^2\theta e^{-2r}+\sin^2\theta e^{2r} & -\cos\theta\sin\theta e^{-2r}+\cos\theta\sin\theta e^{2r} \\ & \\
-\cos\theta\sin\theta e^{-2r}+\cos\theta\sin\theta e^{2r} & \sin^2\theta e^{-2r}+\cos^2\theta e^{2r} \end{array}\right)\right| \\
&=(2\bar{N}+1)^2\left(\cos^2\theta\sin^2\theta e^{-4r}+\cos^4\theta+\sin^4\theta+\sin^2\theta\cos^2\theta e^{4r} \right. \nonumber\\
&\;\;\;\;\left.-\cos^2\theta\sin^2\theta e^{-4r}+\cos^2\theta\sin^2\theta+\sin^2\theta\cos^2\theta-\cos^2\theta\sin^2\theta e^{4r}\right) \\
&=(2\bar{N}+1)^2(\cos^2\theta+\sin^2\theta)^2 \\
&=(2\bar{N}+1)^2. \label{eq:gCvM}
\end{align}
For example, the determinant of a covariance matrix for the squeezed state with zero-mean (fix $\theta=0$) is determined by
\begin{equation}
V(\hat{\rho}_\tn{sq}):=V_\tn{sq}=(2\bar{N}+1)
\left( \begin{array}{cc}
e^{-2r} & 0 \\ & \\
0 & e^{2r} \end{array}\right) \;\;\tn{and}\;\;
\det V_\tn{sq}=(2\bar{N}+1)^2.
\end{equation}

\section{Information capacity for the bosonic Gaussian-noise channel}
For a single-mode bosonic Gaussian state $\hat{\rho}$ wth an input mean photon number $\bar{N}$, the noiseless channel capacity (in nats)~\cite{sYO93,sCD94} is given by
\begin{equation}
C(\cl{N}_0,\bar{N})=g(\bar{N}):=(\bar{N}+1)\log(\bar{N}+1)-\bar{N}\log\bar{N}.
\end{equation}
Also it was known that a pure-lossy channel's capacity (in nats) for a single-mode Gaussian channel $\cl{N}_{\tau,0}$ wth the input mean photon number $\bar{N}$ is as follows:
\begin{equation}
C(\cl{N}_{\tau,0},\bar{N})=g(\tau\bar{N}).
\end{equation}
Note that, in the usual case of the Gaussian information capacity for the thermal-noise channel $C(\cl{N}_{\tau,\bar{N}_E},\bar{N})$, it is bounded by
\begin{equation}
\chi(\cl{N}_{\tau,\bar{N}_E},\bar{N})\le C(\cl{N}_{\tau,\bar{N}_E},\bar{N})\le C_{\max}(\cl{N}_{\tau,\bar{N}_E},\bar{N}),
\end{equation}
where $\chi(\cdot)$ denotes the Holevo capacity and $C_{\max}(\cdot)$ arbitrary maximal information capacity for the thermal-noise channel. ($\bar{N}_E$ denotes the mean photon number for the environmental noise.)

Now we know that the Holevo capacity with coherent-state encoding~\cite{sGLMS03} is given in the form of
\begin{align}
\chi(\cl{N}_{\tau,\bar{N}_E},\bar{N})&=\max_{\bar{\alpha}}S\left(\cl{N}_{\tau,\bar{N}_E}\left(\int\frac{e^{-\frac{|\alpha|^2}{\bar{N}}}}{\pi\bar{N}}\proj{\alpha}{\alpha}d\alpha\right)\right)-\int\frac{e^{-\frac{|\alpha|^2}{\bar{N}}}}{\pi\bar{N}}S\left(\cl{N}_{\tau,\bar{N}_E}(\proj{\alpha}{\alpha})\right)d\alpha \\
&=g(\tau\bar{N}+(1-\tau)\bar{N}_E)-g((1-\tau)\bar{N}_E),
\end{align}
which describes the classical capacity on the Gaussian thermal-noise channel. The maximal information capacity (i.e., the upper bound) is as follows: Let $\bar{\rho}=\sum_{i}p_i\hat{\rho}_i$ be the average state of an input ensemble $\{p_i,\hat{\rho}_i\}$ subject to the mean photon number constraint $\bar{N}$. Then, we can find the upper bound on the thermal-noise channel as
\begin{align}
C(\cl{N}_{\tau,\bar{N}_E},\bar{N})
&:=\lim_{n\to\infty}\frac{1}{n}\chi(\cl{N}_{\tau,\bar{N}_E}^{\otimes n},n\bar{N}) \\
&=\lim_{n\to\infty}\max_{\bar{\rho}=\{p_i,\hat{\rho}_i\}}\frac{1}{n}\left[S\left(\cl{N}_{\tau,\bar{N}_E}^{\otimes n}(\bar{\rho}_n)\right)-\sum_ip_iS\left(\cl{N}_{\tau,\bar{N}_E}^{\otimes n}(\hat{\rho}^i_n)\right)\right] \\
&\le\lim_{n\to\infty}\max_{\bar{\rho}}\frac{1}{n}S\left(\cl{N}_{\tau,\bar{N}_E}^{\otimes n}(\bar{\rho}_n)\right)-\lim_{n\to\infty}\min_{\hat{\rho}}\frac{1}{n}S\left(\cl{N}_{\tau,\bar{N}_E}^{\otimes n}(\hat{\rho}_n)\right) \\
&\le\max_{\bar{\rho}}S\left(\cl{N}_{\tau,\bar{N}_E}(\bar{\rho})\right)-\lim_{n\to\infty}\min_{\hat{\rho}}\frac{1}{n}S\left(\cl{N}_{\tau,\bar{N}_E}^{\otimes n}(\hat{\rho}_n)\right) \\
&=g(\tau\bar{N}+(1-\tau)\bar{N}_E)-\lim_{n\to\infty}\min_{\hat{\rho}}\frac{1}{n}S\left(\cl{N}_{\tau,\bar{N}_E}^{\otimes n}(\hat{\rho}_n)\right) \\
&:=C_{\max}(\cl{N}_{\tau,\bar{N}_E},\bar{N}),
\end{align}
where the first inequality comes from minimization of the von Neumann entropy, the second inequality from the subadditivity of the entropy, and the final one from the fact that Gaussian states also maximize the entropy~\cite{sWGC06}.

For the mean photon number constraint $\bar{N}$, $\T(\hat{H}\bar{\rho})\le2\bar{N}+1$ where the Hamiltonian for the quantum harmonic oscillator is given by $\hat{H}=\frac{1}{2}[\hat{p}^2+\hat{q}^2]$. Note that the number operator $\hat{N}=\frac{\hat{H}-1}{2}$. In this case, the Gaussian minimum output entropy conjecture~\cite{sGGLMS04,sG04} has the form of
\begin{equation}
\lim_{n\to\infty}\min_{\hat{\rho}}\frac{1}{n}S\left(\cl{N}_{\tau,\bar{N}_E}^{\otimes n}(\hat{\rho}_n)\right)=g((1-\tau)\bar{N}_E).\;\;\;\tn{[C]}
\end{equation}
If above conjecture [C] is true, then $\chi(\cl{N}_{\tau,\bar{N}_E},\bar{N})= C(\cl{N}_{\tau,\bar{N}_E},\bar{N})= C_{\max}(\cl{N}_{\tau,\bar{N}_E},\bar{N})$. Recently, it was proved its equality for a specific case in Refs.~\cite{sGGCH14,sMGH14}, i.e., there exists a Gaussian state $\hat{\rho}$ attaining its minimum value.

Our main goal is to constructing a universal formula on the information capacity for the zero-mean Gaussian state $\hat{\rho}$ given by its covariance matrix as
\begin{equation}
V_G=(2\bar{N}_\tn{th}+1)O(\theta)T(2r)O^\top(\theta),
\end{equation}
where $\bar{N}_\tn{th}$ is the thermal photon number of the general Gaussian-noise mode $\hat{b}$. Following Lemma is crucial in the proof of the next theorem.

\begin{lemma}[Determinant of covariance matrices] 
For a covariance matrix for general Gaussian states with the mean photon number $\bar{N}$, we have
\begin{equation}
\det{V_G}=\det{V_\tn{th}}=(2\bar{N}+1)^2.
\end{equation}
\end{lemma}

\emph{Proof}. By using the representation of the general CvM $V_G$, the proof is straightforward (See Eq.~(\ref{eq:gCvM}) in Appendix~\ref{app:cov}). \qed

\begin{theorem}[Strong upper bound for Gaussian information capacity] 
\label{thm:mainbd}
Let $\cl{N}_{\tau,\langle V_G\rangle}$ be a general Gaussian-noise channel with its input Gaussian state $\hat{\rho}$ having the mean photon number $\bar{N}$. Then the upper bound of the information capacity for $\cl{N}_{\tau,\langle V_G\rangle}$ is given by ($\forall\tau\in[0,1]$)
\begin{equation}
C(\cl{N}_{\tau,\langle V_G\rangle},\bar{N})\le g\left(\tau\bar{N}+(1-\tau)\bar{N}_E\right)-(1-\tau)g\left((\sqrt{{\det V_G}}-1)/2\right),
\end{equation}
where $\langle\hat{a}^\dag\hat{a}\rangle=\bar{N}$ is the mean photon number for the input bosonic mode $\hat{a}$, and $\langle\hat{b}^\dag\hat{b}\rangle=\bar{N}_E:=\langle V_G\rangle$.
\end{theorem}

\emph{Proof}. Let $C_{\max}(\cl{N}_{\tau,V_G},\bar{N})$ be the maximal value for the regularized channel capacity for general Gaussian-noise channel with zero-mean: Then there exists Gaussian state $\hat{\rho}\in\{p_i,\hat{\rho}_i\}$ such that
\begin{align}
C(\cl{N}_{\tau,V_G},\bar{N})&\le C_{\max}(\cl{N}_{\tau,V_G},\bar{N}) \\
&= g\left(\tau\bar{N}+(1-\tau)\bar{N}_E\right)-\lim_{n\to\infty}\min_{\hat{\rho}}\frac{1}{n}S\left(\cl{N}_{\tau,V_G}^{\otimes n}(\hat{\rho}_n)\right).
\end{align}
Now we assume that $\hat{\rho}_\tn{noise}$ has a covariance matrix $V_G$. Then, by using the QEPI~\cite{sKS14,sPMG14}, we have
\begin{align}
S\left(\cl{N}_{\tau,V_G}^{\otimes n}(\hat{\rho}_n)\right) 
&\ge\tau S(\hat{\rho}_n)+(1-\tau)S(\hat{\rho}_\tn{noise}^{\otimes n}) \label{eq:qepi} \\
&=\tau S(\hat{\rho}_n)+n(1-\tau) S(\hat{\rho}_\tn{noise}) \label{eq:subadd} \\
&\ge n(1-\tau) S(\hat{\rho}_\tn{noise}) \label{eq:positivity} \\
&=n(1-\tau)g(\bar{N}_E) \label{eq:def} \\
&:=n(1-\tau)g\left((\sqrt{{\det V_G}}-1)/2\right),
\end{align}
where Eq.~(\ref{eq:qepi}) comes from the EPI for Gaussian quantum states, Eq.~(\ref{eq:subadd}) from the subadditivity of the von Neumann entropy for separable environments, Eq.~(\ref{eq:positivity}) from the positivity of the entropy, and finally Eq.~(\ref{eq:def}) is given by the definition. Thus, we have
\begin{equation}
C(\cl{N}_{\tau,\langle V_G\rangle},\bar{N})\le g\left(\tau\bar{N}+(1-\tau)\bar{N}_E\right)-(1-\tau)g\left((\sqrt{{\det V_G}}-1)/2\right).
\end{equation}
This completes the proof. \qed

\begin{remark}[Upper bound on thermal-noise channel]
Let $\cl{N}_{\tau,\langle V_\tn{th}\rangle}$ be the bosonic Gaussian thermal-noise channel with the mean photon numbers $\bar{N}$ and $\bar{N}_E$ for the input-mode and environmental-mode, respectively. Then we have
\begin{equation}
C(\cl{N}_{\tau,\langle V_\tn{th}\rangle},\bar{N})\le g\left(\tau\bar{N}+(1-\tau)\bar{N}_E\right)-(1-\tau)g\left(\bar{N}_E\right).
\end{equation}
\end{remark}
Since $\tfrac{1}{2}(\sqrt{{\det V_{\tn{th}}}}-1)=\tfrac{1}{2}(2\bar{N}_E+1)-\tfrac{1}{2}=\bar{N}_E$ and $\bar{N}_E=\bar{N}_\tn{th}$ for the thermal noise channel. Our result (Theorem~\ref{thm:mainbd}) is conceptually including the K\"{o}nig and Smith's previous work~\cite{sKS13}.

%


\begin{thebibliography}{47}%
%
\bibitem{S48}
C. E. Shannon,
``A mathematical theory of communication,"
\href{http://dx.doi.org/10.1002/j.1538-7305.1948.tb01338.x}{
Bell Syst. Tech. J.~\textbf{27}, 379--423, 623--656 (1948)}.
%
\bibitem{NC00}
M. A. Nielsen and I. L. Chuang, 
``Quantum Computation and Quantum Information," 
Cambridge University Press (2000).
%
\bibitem{W13}
M. M. Wilde, 
``Quantum Information Theory," 
Cambridge University Press (2013).
%
\bibitem{H06}
A. S. Holevo, 
``The additivity problem in quantum information theory,"
In Proceedings of the International Congress of Mathematicians, (Madrid, Spain, 2006).
%
\bibitem{HW01} 
A. S. Holevo and R. F. Werner,
``Evaluating capacities of bosonic Gaussian channels,"
\href{http://dx.doi.org/10.1103/PhysRevA.63.032312}{
\pra~\textbf{63}, 032312 (2001)}.
%
\bibitem{H09}
M. B. Hastings, 
``Superadditivity of communication capacity using entangled inputs," 
\href{http://dx.doi.org/10.1038/nphys1224}{
Nat. Phys.~\textbf{5}, 255--257 (2009)}.
%
\bibitem{SY08}
G. Smith and J. Yard, 
``Quantum Communication with Zero-Capacity Channels," 
\href{http://dx.doi.org/10.1126/science.1162242}{
Science~\textbf{321}, 1812--1815 (2008)}.
%
\bibitem{LWZG09}
K. Li, A. Winter, X. Zou, and G. Guo, 
``Private Capacity of Quantum Channels is Not Additive," 
\href{https://doi.org/10.1103/PhysRevLett.103.120501}{
\prl~\textbf{103}, 120501 (2009)}.
%
\bibitem{GGCH14}
V. Giovannetti, R. Garc\'{i}a-Patr\'{o}n, N. J. Cerf and A. S. Holevo,
``Ultimate classical communication rates of quantum optical channels," 
\href{http://dx.doi.org/10.1038/nphoton.2014.216}{
Nat. Photonics~\textbf{8}, 796--800 (2014)}.
%
\bibitem{MGH14}
A. Mari, V. Giovannetti, and A. S. Holevo,
``Quantum state majorization at the output of bosonic Gaussian channels," 
\href{http://dx.doi.org/10.1038/ncomms4826}{
Nat. Commun.~\textbf{5}, 3826 (2014)}.
%
\bibitem{KS13}
R. K\"{o}nig and G. Smith,
``Limits on classical communication from quantum entropy power inequalities," 
\href{http://dx.doi.org/10.1038/nphoton.2012.342}{
Nat. Photonics~\textbf{7}, 142--146 (2013)}.
%
\bibitem{KS13+} 
R. K\"{o}nig and G. Smith,
``Classical Capacity of Quantum Thermal Noise Channels to within 1.45 Bits,"
\href{http://dx.doi.org/10.1103/PhysRevLett.110.040501}{
\prl~\textbf{110}, 040501 (2013)}.
%
\bibitem{KS14}
R. K\"{o}nig and G. Smith,
``The Entropy Power Inequality for Quantum Systems," 
\href{http://dx.doi.org/10.1109/TIT.2014.2298436}{
IEEE Trans. Inf. Theory~\textbf{60}, 1536--1548 (2014)}.
%
\bibitem{L78}
E. H. Lieb,
``Proof of an entropy conjecture of Wehrl," 
\href{http://dx.doi.org/10.1007/BF01940328}{
Commun. Math. Phys.~\textbf{62}, 35--41 (1978)}.
%
\bibitem{DCT91}
A. Dembo, T. Cover, and J. A. Thomas,
``Information theoretic inequalities," 
\href{http://dx.doi.org/10.1109/18.104312}{
IEEE Trans. Inf. Theory~\textbf{37}, 1501--1518 (1991)}.
%
\bibitem{B75}
W. Beckner,
``Inequalities in Fourier analysis," 
\href{http://dx.doi.org/10.2307/1970980}{
Ann. Math.~\textbf{102}, 159--182 (1975)}.
%
\bibitem{BL76}
H. J. Brascamp and E. H. Lieb,
``Best constants in Young's inequality, its converse, and its generalization to more than three functions," 
\href{http://dx.doi.org/10.1016/0001-8708(76)90184-5}{
Adv. Math.~\textbf{102}, 151--172 (1976)}.
%
\bibitem{R11}
O. Rioul,
``Information theoretic proofs of entropy power inequalities," 
\href{http://dx.doi.org/10.1109/TIT.2010.2090193}{
IEEE Trans. Inf. Theory~\textbf{57}, 33--55 (2011)}.
%
\bibitem{S59}
A. J. Stam,
``Some inequalities satisfied by the quantities of information of Fisher and Shannon," 
\href{http://dx.doi.org/10.1016/S0019-9958(59)90348-1}{
Inf. Control~\textbf{2}, 101--112 (1959)}.
%
\bibitem{B65}
 N. M. Blachman,
``The convolution inequality for entropy powers," 
\href{http://dx.doi.org/10.1109/TIT.1965.1053768}{
IEEE Trans. Inf. Theory~\textbf{11}, 267--271 (1965)}.
%
\bibitem{PMG14}
G. de Palma, A. Mari, and V. Giovannetti, 
``A generalization of the entropy power inequality to bosonic quantum systems," 
\href{http://dx.doi.org/10.1038/nphoton.2014.252}{
Nat. Photonics~\textbf{8}, 958--964 (2014)}.
%
\bibitem{PMLG15}
G. de Palma, A. Mari, S. Lloyd, and V. Giovannetti, 
``Multimode quantum entropy power inequality," 
\href{http://dx.doi.org/10.1103/PhysRevA.91.032320}{
\pra~\textbf{91}, 032320 (2015)}.
%
\bibitem{ADO16}
K. Audenaert, N. Datta, and M. Ozols, 
``Entropy power inequalities for qudits," 
\href{http://dx.doi.org/10.1063/1.4950785}{
J. Math. Phys.~\textbf{57}, 052202 (2016)}.
%
\bibitem{K15}
R. Koenig, 
``The conditional entropy power inequality for Gaussian quantum states," 
\href{http://dx.doi.org/10.1063/1.4906925}{
J. Math. Phys.~\textbf{56}, 022201 (2015)}.
%
\bibitem{JLJ18}
K. Jeong, S. Lee, and H. Jeong, 
``Conditional quantum entropy power inequality for $d$-level quantum systems," 
\href{http://dx.doi.org/10.1088/1751-8121/aab037}{
J. Phys. A: Math. Theor.~\textbf{51}, 145303 (2018)}.
%
\bibitem{CLL16}
E. A. Carlen, E. H. Lieb, and M. Loss, 
``On a quantum entropy power inequality of Audenaert, Datta, and Ozols," 
\href{http://dx.doi.org/10.1063/1.4953638}{
J. Math. Phys.~\textbf{57}, 062203 (2016)}.
%
\bibitem{PT17}
G. de Palma and D. Trevisan, 
``The Conditional Entropy Power Inequality for Bosonic Quantum Systems," 
\href{https://doi.org/10.1007/s00220-017-3082-8}{
Commun. Math. Phys.~\textbf{360}, 639--662 (2018)}.
%
\bibitem{PMG14+} 
G. de Palma, A. Mari, and V. Giovannetti,
``Classical capacity of Gaussian thermal memory channels,"
\href{http://dx.doi.org/10.1103/PhysRevA.90.042312}{
\pra~\textbf{90}, 042312 (2014)}.
%
\bibitem{K05} 
K. Br\'{a}dler,
``Continuous-variable private quantum channel,"
\href{http://dx.doi.org/10.1103/PhysRevA.72.042313}{
\pra~\textbf{72}, 042313 (2005)}.
%
\bibitem{JKL15}
K. Jeong, J. Kim, and S.-Y. Lee,
``Gaussian private quantum channel with squeezed coherent states,"
\href{http://dx.doi.org/10.1038/srep13974}{
Sci. Rep.~\textbf{5}, 13974 (2015)}.
%
\bibitem{JL16}
K. Jeong and Y. Lim,
``Purification of Gaussian maximally mixed states,"
\href{http://dx.doi.org/10.1016/j.physleta.2016.09.001}{
Phys. Lett. A~\textbf{380}, 3607--3611 (2016)}.
%
\bibitem{H98}
A. S. Holevo,
``The capacity of the quantum channel with general signal states," 
\href{http://dx.doi.org/10.1109/18.651037}{
IEEE Trans. Inf. Theory~\textbf{44}, 269--273 (1998)}.
%
\bibitem{SW97} 
B. Schumacher and M. D. Westmoreland,
``Sending classical information via noisy quantum channels,"
\href{http://dx.doi.org/10.1103/PhysRevA.56.131}{
\pra~\textbf{56}, 131--138 (1997)}.
%
\bibitem{SKPC16}
J. Sch\"{a}fer, E. Karpov, O. V. Pilyavets, and N. J. Cerf, 
``Classical capacity of phase-sensitive Gaussian quantum channels," 
\href{http://arxiv.org/abs/1609.04119}{
arXiv:1609.04119 (2016)}.
%
\bibitem{WP+12} 
C. Weedbrook, S. Pirandola, R. Garc\'{i}a-Patr\'{o}n, N. J. Cerf, T. C. Ralph, J. H. Shapiro, and S. Lloyd,
``Gaussian quantum information,"
\href{http://dx.doi.org/10.1103/RevModPhys.84.621}{
\rmp~\textbf{84}, 621--669 (2012)}.
%
\bibitem{YO93} 
H. P. Yuen and M. Ozawa,
``Ultimate information carrying limit of quantum systems,"
\href{http://dx.doi.org/10.1103/PhysRevLett.70.363}{
\prl~\textbf{70}, 363 (1993)}.
%
\bibitem{CD94} 
C. M. Caves and P. D. Drummond,
``Quantum limits on bosonic communication rates,"
\href{http://dx.doi.org/10.1103/RevModPhys.66.481}{
\rmp~\textbf{66}, 481--537 (1994)}.
%
\bibitem{GLMS03} 
V. Giovannetti, S. Lloyd, L. Maccone, and P. W. Shor,
``Broadband channel capacities,"
\href{http://dx.doi.org/10.1103/PhysRevA.68.062323}{
\pra~\textbf{68}, 062323 (2003)}.
%
\bibitem{GGLMS04} 
V. Giovannetti, S. Guha, S. Lloyd, L. Maccone, and J. H. Shapiro,
``Minimal output entropy of bosonic channels: a conjecture,"
\href{http://dx.doi.org/10.1103/PhysRevA.70.032315}{
\pra~\textbf{70}, 032315 (2004)}.
%
\bibitem{GGL+04} 
V. Giovannetti, S. Guha, S. Lloyd, L. Maccone, J. H. Shapiro, and H. P. Yuen,
``Classical Capacity of the Lossy Bosonic Channel: The Exact Solution,"
\href{http://dx.doi.org/10.1103/PhysRevLett.92.027902}{
\prl~\textbf{92}, 027902 (2004)}.
%
\bibitem{WGC06} 
M. M. Wolf, G. Giedke, and J. I. Cirac,
``Extremality of Gaussian Quantum States,"
\href{http://dx.doi.org/10.1103/PhysRevLett.96.080502}{
\prl~\textbf{96}, 080502 (2006)}.
%
\bibitem{PLM12}
O. V. Pillyavets, C. Lupo, and S. Mancini,
``Methods for Estimating Capacities and Rates of Gaussian Quantum Channels,''
\href{http://dx.doi.org/10.1109/TIT.2012.2191475}{
IEEE Trans. Inf. Theory~\textbf{58}, 6126--6164 (2012)}.
%
\bibitem{G04}
S. Guha, 
``Classical capacity of the free-space quantum-optical channel," 
S. M. Thesis, MIT (2004).
%
\bibitem{GES08}
S. Guha, B. I. Erkmen, and J. H. Shapiro,
``The entropy photon-number inequality and its consequences,"
\href{http://dx.doi.org/10.1109/ITA.2008.4601037}{
Information Theory and Applications Workshop, pp. 128--130 (2008)}; 
\href{http://arxiv.org/abs/0710.5666}{arXiv:0710.5666}.
%
\bibitem{HK18} 
S. Huber and R. K\"{o}nig,
``Coherent state coding approaches the capacity of non-Gaussian bosonic noise channels,"
\href{http://dx.doi.org/10.1088/1751-8121/aab7ff}{
J. Phys. A: Math. Theor.~\textbf{51}, 184001 (2018)}.

\end{thebibliography}

\begin{thebibliography}{7}
%
\bibitem{sYO93} 
H. P. Yuen and M. Ozawa,
``Ultimate information carrying limit of quantum systems,"
\href{http://dx.doi.org/10.1103/PhysRevLett.70.363}{
\prl~\textbf{70}, 363 (1993)}.
%
\bibitem{sCD94} 
C. M. Caves and P. D. Drummond,
``Quantum limits on bosonic communication rates,"
\href{http://dx.doi.org/10.1103/RevModPhys.66.481}{
\rmp~\textbf{66}, 481--537 (1994)}.
%
\bibitem{sGLMS03} 
V. Giovannetti, S. Lloyd, L. Maccone, and P. W. Shor,
``Broadband channel capacities,"
\href{http://dx.doi.org/10.1103/PhysRevA.68.062323}{
\pra~\textbf{68}, 062323 (2003)}.
%
\bibitem{sWGC06} 
M. M. Wolf, G. Giedke, and J. I. Cirac,
``Extremality of Gaussian Quantum States,"
\href{http://dx.doi.org/10.1103/PhysRevLett.96.080502}{
\prl~\textbf{96}, 080502 (2006)}.
%
\bibitem{sGGLMS04} 
V. Giovannetti, S. Guha, S. Lloyd, L. Maccone, and J. H. Shapiro,
``Minimal output entropy of bosonic channels: a conjecture,"
\href{http://dx.doi.org/10.1103/PhysRevA.70.032315}{
\pra~\textbf{70}, 032315 (2004)}.
%
\bibitem{sG04}
S. Guha, 
``Classical capacity of the free-space quantum-optical channel," 
S. M. Thesis, MIT (2004).
%
\bibitem{sGGCH14}
V. Giovannetti, R. Garc\'{i}a-Patr\'{o}n, N. J. Cerf and A. S. Holevo,
``Ultimate classical communication rates of quantum optical channels," 
\href{http://dx.doi.org/10.1038/nphoton.2014.216}{
Nat. Photonics~\textbf{8}, 796--800 (2014)}.
%
\bibitem{sMGH14}
A. Mari, V. Giovannetti, and A. S. Holevo,
``Quantum state majorization at the output of bosonic Gaussian channels," 
\href{http://dx.doi.org/10.1038/ncomms4826}{
Nat. Commun.~\textbf{5}, 3826 (2014)}.
%
\bibitem{sKS14}
R. K\"{o}nig and G. Smith,
``The Entropy Power Inequality for Quantum Systems," 
\href{http://dx.doi.org/10.1109/TIT.2014.2298436}{
IEEE Trans. Inf. Theory~\textbf{60}, 1536--1548 (2014)}.
%
\bibitem{sPMG14}
G. de Palma, A. Mari, and V. Giovannetti, 
``A generalization of the entropy power inequality to bosonic quantum systems," 
\href{http://dx.doi.org/10.1038/nphoton.2014.252}{
Nat. Photonics~\textbf{8}, 958--964 (2014)}.
%
\bibitem{sKS13}
R. K\"{o}nig and G. Smith,
``Limits on classical communication from quantum entropy power inequalities," 
\href{http://dx.doi.org/10.1038/nphoton.2012.342}{
Nat. Photonics~\textbf{7}, 142--146 (2013)}.




\end{thebibliography}
\end{document}